\documentclass[12pt]{article}
\usepackage{latexsym,epsfig,graphicx,a4wide,amssymb}

\def\de{\partial}
\def\a{\alpha}
\def\b{\beta}
\def\g{\gamma}

\def\d{\delta}

{\rm }
\def\la{\lambda}
\def\La{\Lambda}
\def\k{\kappa}
\def\m{\mu}
\def\n{\nu}
\def\r{\rho}

\def\s{\sigma}

\def\th{\theta}
\def\z{\zeta}
\def\x{\chi}
\def\be{\begin{equation}}
 \def\ee{\end{equation}}
 \def\bea{\begin{eqnarray}}
 \def\eea{\end{eqnarray}}
 \def\a{\alpha}
 \def\b{\beta}
 \def\g{\gamma}
 \def\d{\delta}
 
 \def\s{\sigma}

\def\L{\Lambda}

\def\2{\frac{1}{2}}
\def\4{\frac{1}{4}}

\begin{document}

\begin{titlepage}

\begin{centering}
\vspace{1cm}
{\Large {\bf Black Holes and Black String-like Solutions in
Codimension-2 Braneworlds $^*$}}\\

\vspace{1.5cm}

 {\bf Eleftherios Papantonopoulos} \\
\vspace{.2in}

 Department of Physics, National Technical University of
Athens, \\
Zografou Campus GR 157 73, Athens, Greece. \\
e-mail address: lpapa@central.ntua.gr\\

\end{centering}
\vspace{4cm}

\begin{abstract}

We discuss black hole solutions  with a Gauss-Bonnet term in the
bulk and an induced gravity term on a thin  brane of
codimension-2. We show that these black holes can be localized on
the brane, and they can  be extended further into the bulk by a
warp function. These solutions have regular horizons and no other
curvature singularities appear apart from the string-like ones.
The projection of the Gauss-Bonnet term on the brane imposes a
constraint relation  which dictates the form of matter on the
brane and in the bulk.

\end{abstract}

\vspace{6.5cm}
\begin{flushleft}
$^{*}~$Plenary talk given at the 7th Friedmann International
Seminar on Gravitation and Cosmology, 29 June-5 July 2008,
Jo\~{a}o Pessoa, Brazil.

\end{flushleft}

\end{titlepage}

\section{Introduction}

Recently there has been some interest in codimension-2
braneworlds. The most attractive feature of these models is that
the vacuum energy (tension) of the brane
 instead of curving the brane world-volume, merely induces a deficit angle in the
 bulk solution around the brane. This observation
led  several people to utilize this property in order to self-tune
the effective cosmological constant  to zero and provide a
solution to the cosmological constant problem \cite{6d}. However,
soon it was realized \cite{Cline}
   that one can only find nonsingular
solutions if the brane energy momentum tensor is proportional to
its induced metric. To reproduce an effective four-dimensional
Einstein equation on the brane one has to introduce a cut-off
(brane thickness) \cite{Kanno:2004nr,Vinet:2004bk,Navarro:2004di}
with the price of loosing the predictability of the theory.
Alternatively, in the thin brane limit four dimensional gravity is
recovered as the dynamics of the induced metric on the brane if
the gravitational action is modified by the inclusion of either a
Gauss-Bonnet term~\cite{Bostock:2003cv} or an induced gravity term
on the brane~\cite{Papantonopoulos:2005ma}.

We are still lacking an understanding of time dependent
cosmological solutions in codimension-2 braneworlds. In the thin
brane limit, because the energy momentum tensors on the brane and
in the bulk are related, the brane equation of state and energy
density are tuned and we cannot get the standard cosmology on the
brane~\cite{Kofinas:2005py,Papantonopoulos:2005nw}. One then has
to regularize the codimension-2 branes by introducing some
thickness and then consider matter on them
\cite{regular,PST,ppz,tas}. To have a cosmological evolution on
the regularized branes the brane world-volume should be expanding
and in general
 the bulk space should also evolve in time. This is a formidable task,
  so an alternatively approach was followed in~\cite{Papantonopoulos:2007fk,Minamitsuji:2007fx}
by considering a codimension-1 brane moving in the regularized
static background. The resulting cosmology, however, was
unrealistic having a negative Newton's constant (for a review on
the cosmology in six dimensions
see~\cite{Papantonopoulos:2006uj}).

We do not either fully understand black hole solutions on
codimension-2 braneworlds. Recently a six-dimensional black hole
localized on a 3-brane of codimension-2~\cite{Kaloper:2006ek} was
proposed. These solutions are generalization of the $4D$ Aryal,
Ford, Vilenkin~\cite{Aryal:1986sz,Achucarro:1995nu} black hole
pierced by a cosmic string adjusted to the codimension-2 branes
with a conical structure in the bulk and deformations
 accommodating the deficit angle. However, it is not
clear how to realize these solutions in the thin brane limit where
high curvature terms are needed to accommodate matter on the
brane. Generalizations to include rotations were presented
in~\cite{Kiley:2007wb}.

The  localization of a black hole on the brane and its extension
to the bulk is a difficult task. In codimension-1 braneworlds the
first attempt was to consider the Schwarzschild metric and study
its black string extension into the bulk~\cite{Chamblin:1999by}.
Unfortunately, as suspected by the authors, this string is
unstable to classical linear perturbations~\cite{BSINS} (for a
recent review see~\cite{Harmark:2007md}). Since then, several
authors have attempted to find the full metric using numerical
techniques \cite{BHNUM}. Analytically, the brane metric equations
of motion were considered with the only bulk input coming from the
projection of the Weyl tensor~\cite{SMS} onto the brane. Since
this system is not closed because it contains an unknown bulk
dependent term, assumptions have to be made either in the form of
the metric or on the Weyl term~\cite{BBH}.

 A lower dimensional version of a black hole living
on a (2+1)-dimensional braneworld was considered in~\cite{EHM} by
Emparan, Horowitz, and Myers. They based their analysis on the
so-called C-metric~\cite{Kinnersley:zw} modified by a cosmological
constant term. They found a BTZ black hole~\cite{Banados:1992wn}
on the brane which can be extended as a BTZ black string in a
four-dimensional AdS bulk. Their thermodynamical stability
analysis showed that the black string remains a stable
configuration when its transverse size is comparable to the
four-dimensional AdS radius, being destabilized by the
Gregory-Laflamme instability~\cite{BSINS} above that scale,
breaking up to a BTZ black hole on a 2-brane.

In this talk we will discuss black holes on a thin conical brane
and their extension into a five and six-dimensional bulk with a
Gauss-Bonnet
term~\cite{CuadrosMelgar:2007jx,CuadrosMelgar:2008kn}. In the case
of a five-dimensional bulk~\cite{CuadrosMelgar:2007jx} we had
found that the BTZ black hole and its short distance
corrections~\cite{Zanelli1996} solve the junction conditions on a
conical 2-brane. These solutions in the bulk are BTZ string-like
objects with regular horizons and no pathologies. The warping to
five dimensions depends on the length $\sqrt{\alpha}$ where
$\alpha$ is the Gauss-Bonnet coupling, and this length scale
defines the shape of the horizon. Consistency of the bulk
solutions requires a fine-tuned relation between the Gauss-Bonnet
coupling and the five-dimensional cosmological constant.


In the case of six-dimensional bulk, we found solutions of
four-dimensional Schwarzschild-AdS black holes on a 3-brane which
in the six-dimensional spacetime look like black string-like
objects with regular horizons~\cite{CuadrosMelgar:2008kn}. In the
case of constant deficit angle the localization of the
four-dimensional black hole requires matter in the two extra
dimensions. The energy-momentum tensor corresponding to this
matter scales as $1/r^6$.  This fact defines a length scale in the
six-dimensional spacetime above which we recover the standard
four-dimensional General Relativity (GR), while at small distances
GR is strongly modified.

\section{BTZ String-Like Solutions in Five-Dimensional\\ Braneworlds of Codimension-2}

We consider the following gravitational action in five dimensions
with a Gauss-Bonnet term in the bulk and an induced
three-dimensional curvature term on the brane
\begin{eqnarray}\label{AcGBIG}
S_{\rm grav}&=&\frac{M^{3}_{5}}{2}\left\{ \int d^5 x\sqrt{-
g}\left[ R
+\alpha\left( R^{2}-4 R_{MN}R^{MN}+
R_{MNKL}R^{MNKL}\right)\right] \right.\nonumber\\
&+& \left. r^{2}_{c} \int d^3x\sqrt{- g^{(3)}}\,R^{(3)}
\right\}+\int d^5 x \mathcal{L}_{bulk}+\int d^3 x
\mathcal{L}_{brane}\,,\label{5daction}
\end{eqnarray}
where $\alpha\, (\geq0)$ is the GB coupling constant and
$r_c=M_{3}/M_{5}^3$ is the induced gravity ``cross-over" scale.
The bulk metric is \be ds_5^2=g_{\m\n}(x,\rho)dx^\m
dx^\n+a^{2}(x,\rho)d\rho^2+L^2(x,\rho)d\th^2~,\label{5dmetric} \ee
where $g_{\mu\nu}(x,0)$ is the braneworld metric and $x^{\mu}$
denote three  dimensions, $\mu,\nu=0,1,2$~ whereas $\rho,\th$
denote the radial and angular coordinates of the two extra
dimensions.

The Einstein equations resulting from the variation of the
action~(\ref{5daction}) are \be
 G^{(5)N}_M + r_c^2
G^{(3)\n}_\m g_M^\m g^N_\n {\d(\rho) \over 2 \pi L}-\alpha
H_{M}^{N} =\frac{1}{M^{3}_{5}} \left[T^{(B)N}_M+T^{(br)\n}_\m
g_M^\m g^N_\n {\d(\rho) \over 2 \pi L}\right]~, \label{einsequat3}
\ee where $H_M^N$ is coming from the Gauss-Bonnet term.
 To obtain the braneworld
equations we expand the metric around the brane as $
L(x,\rho)=\beta(x)\rho+O(\rho^{2})~.$  We  demand that the space
in the vicinity of the conical singularity is regular which
imposes the supplementary conditions that $\de_\m \b=0$ and
$\partial_{\rho}g_{\mu\nu}(x,0)=0$~\cite{Bostock:2003cv}.

The extrinsic curvature  is given by $K_{\m\n}=g'_{\m\n}$.
 The second derivatives of the metric
functions contain $\d$-function singularities  \bea
{L'' \over L}&=&-(1-L'){\d(\rho) \over L}+ {\rm non-singular~terms}~,\\
{K'_{\m\n} \over L}&=&K_{\m\n}{\d(\rho) \over L}+ {\rm
non-singular~terms}~. \eea

From the above singularity expressions and using the Gauss-Codazzi
equations, we can  match the singular parts of the Einstein
equations (\ref{einsequat3}) and get the following ``boundary"
Einstein equations
 \be G^{(3)}_{\m\n}={1 \over M_{(5)}^3 (r_c^2+8\pi
(1-\b)\a)}T^{(br)}_{\m\n}+{2\pi (1-\b) \over r_c^2+8\pi
(1-\b)\a}g_{\m\n} \label{einsteincomb3}~. \ee

We assume that there is a localized (2+1) black hole on the brane.
The brane metric is \be
ds_{3}^{2}=\left(-n(r)^{2}dt^{2}+n(r)^{-2}dr^{2}+r^{2}d\phi^{2}\right)~.
\label{3dmetric}\ee
 We will look for black string solutions of the
 Einstein equations~(\ref{einsequat3}) using the
 five-dimensional metric~(\ref{5dmetric}) in the form
\be ds_5^2=f^{2}(\rho)\left(-n(r)^{2}dt^{2}+n(r)^{-2}dr^{2}+r^{2}
d\phi^{2}\right)+a^{2}(r,\rho)d\rho^2+L^2(r,\rho)d\th^2~.\label{5smetricc}
\ee

 The space outside the
 conical singularity is regular, therefore, we  demand that the warp function $ f(\rho) $ is
  also regular
 everywhere. We assume that there is only a cosmological constant $\Lambda_{5}$
in the bulk and we take $a(r,\rho)=1$. Then, from the bulk
Einstein equations \be G^{(5)}_{MN}-\alpha
H_{MN}=-\frac{\Lambda_{5}}{M^{3}_{5}}g_{MN}~,\ee

we find the solutions which are summarized in Table
1~\cite{CuadrosMelgar:2007jx}.

\begin{table}[here]
\begin{center}
\begin{tabular}{|c|c|c|c|c|c|}
  \hline
  $n(r)$ & $f(\r)$ & $L(\r)$ & $-\L_5$ & Constraints \\
  \hline
  BTZ & $\cosh\left(\frac{\r}{2\,\sqrt{\a}}\right)$ & $\forall L(\r)$ &
$\frac{3}{4\a}$ &
$L_3^2=4\,\a$ \\
  BTZ & $\cosh\left(\frac{\r}{2\,\sqrt{\a}}\right)$ & $2\,\b\,\sqrt{\a}\,\sinh\left(\frac{\r}{2\,\sqrt{\a}}\right)$ &
  $\frac{3}{4\a}$ & - \\
  BTZ & $\cosh\left(\frac{\r}{2\,\sqrt{\a}}\right)$ & $2\,\b\,\sqrt{\a}\,\sinh\left(\frac{\r}{2\,\sqrt{\a}}\right)$ &
  $\frac{3}{4\a}$ & $L_3^2=4\,\a$ \\
  BTZ & $\pm 1$ & $\frac{1}{\g}\,\sinh\left(\g\,\r\right)$ & $\frac{3}{l^2}$ & $\g=\sqrt{-\frac{2\L_5}{3+4\a\L_5}}$ \\
  $\forall n(r)$ & $\cosh\left(\frac{\r}{2\,\sqrt{\a}}\right)$ & $2\,\b\,\sqrt{\a}\,\sinh\left(\frac{\r}{2\,\sqrt{\a}}\right)$ &
  $\frac{3}{4\a}$ & -  \\
  $\sqrt{-M+\frac{r^2}{L_3^2}-\frac{\z}{r}}$ & $\cosh\left(\frac{\r}{2\,\sqrt{\a}}\right)$ &
$2\,\b\,\sqrt{\a}\,\sinh\left(\frac{\r}{2\,\sqrt{\a}}\right)$ &
$\frac{3}{4\a}$ &
$L_3^2=4\,\a$ \\
  $\sqrt{-M+\frac{r^2}{L_3^2}-\frac{\z}{r}}$ & $\pm 1$ & $2\,\b\,\sqrt{\a}\,\sinh\left(\frac{\r}{2\,\sqrt{\a}}\right)$
  & $\frac{1}{4\a}$ & $\L_5=-\frac{1}{4\a}=-\frac{3}{L_3^2}$ \\
  \hline
\end{tabular}
\end{center}
\caption{BTZ String-Like Solutions in Five-Dimensional Braneworlds
of Codimension-2}\label{table1}
\end{table}

In the above table $L_3$ is the length scale of $AdS_{3}$ space.
Note that in all solutions there is a fine-tuned relation between
the Gauss-Bonnet coupling $\alpha$ and the five-dimensional
cosmological constant $\Lambda_{5}$, except for the solution in
the fourth row.

To introduce a brane we must solve the corresponding junction
conditions given by the Einstein equations on the brane
(\ref{einsteincomb3}) using the induced metric on the brane given
by (\ref{3dmetric}). We found that the BTZ black hole is localized
on the brane in vacuum.

When $n(r)$ is of the form  $n(r)^{2}=-M+r^2/L_3^2-\z/r$,
 which is the BTZ black hole solution with
a short distance correction term and it corresponds to the BTZ
conformally coupled to a scalar field~\cite{Zanelli1996}, the
energy momentum tensor necessary to sustain such a solution on the
brane is given by
 $T_\alpha ^\beta
=\hbox{diag}\left( \zeta/2r^3,\zeta/2r^3,-\zeta/r^3 \right)\, .$

These solutions extend the brane BTZ black hole into the bulk.
Calculating the square of the Riemann tensor we find that at the
AdS horizon ($\rho \rightarrow \infty$) all solutions give finite
result and hence the only singularity is
  the  BTZ black hole singularity extended into the bulk.
  The warp function $f^{2}(\rho)$ gives the shape of a
'throat' to
    the horizon
of the BTZ string-like solution. The size of the horizon is
defined by the scale $\sqrt{\alpha}$ and this scale is fine-tuned
to the length scale of the five-dimensional AdS space.

\section{Black String-Like solutions in Six-Dimensional\\ Braneworlds of Codimension-2 }

 The metric as in the five-dimensional case is
\be ds_6^2=g_{\m\n}(r,\x)dx^\m
dx^\n+a^{2}(r,\x)d\x^2+L^2(r,\x)d\xi^2~,\label{6dmetric} \ee now
with $\mu,\nu=0,1,2,3$ whereas $\x,\xi$ denote the radial and
angular coordinates of the two extra dimensions.

The corresponding Einstein equations are
 \be
 G^{(6)N}_M + r_c^2
G^{(4)\n}_\m g_M^\m g^N_\n {\d(\x) \over 2 \pi L}-\alpha H_{M}^{N}
=\frac{1}{M^{4}_{6}} \left[-\L_{6}+T^{(B)N}_M+T^{(br)\n}_\m g_M^\m
g^N_\n {\d(\x) \over 2 \pi L}\right]~, \label{einsequat} \ee where
$H_{M}^{N}$ is the corresponding six-dimensional term.
 To obtain the braneworld equations we expand the
metric around the 3-brane as $ L(r,\x)=\b(r)\x+O(\x^{2})~,$ and as
in the five-dimensional case the function $L$ behaves as
$L^{\prime}(r,0)=\beta(r)$, where a prime now denotes derivative
with respect to $\x$. The ``boundary" Einstein equations are \bea
G^{(4)}_{\m\n} \left(r_c^2+8\pi (1-\b)\a\right) &=& {1 \over
M_{6}^4}
T^{(br)}_{\m\n}+ 2\pi (1-\b)g_{\m\n} \nonumber \\
&+& \pi L(r,\x)\,E_{\m\n}-2\pi\b\a\,W_{\m\n}
\label{einsteincomb}~, \eea where the term \be E_{\m\n} =
\left(K_{\m\n}-g_{\m\n}\,K\right)~, \label{INDContrib} \ee
 appears because of the presence of the induced
gravity term in the gravitational action, while the term \bea
W_{\m\n} &=& g^{\la\s}\partial_{\x}g_{\m\la}\partial_{\x}g_{\n\s}-
g^{\la\s}\partial_{\x}g_{\la\s}\partial_{\x}g_{\m\n} \nonumber \\
&+&\frac{1}{2}g_{\m\n}\left[\left(g^{\la\s}\partial_{\x}g_{\la\s}\right)^2
-g^{\la\s}g^{\d\r}\partial_{\x}g_{\la\d}\partial_{\x}g_{\s\r}\right]~,
\label{GBContrib} \eea is the Weyl term due to the presence of the
Gauss-Bonnet term in the bulk.

 We will look for black string solutions of the
 Einstein equations~(\ref{einsequat}) using the
 six-dimensional metric~(\ref{6dmetric}) in the form
\bea
ds_6^2&=&F^{2}(\x)\left(-A(r)^{2}dt^{2}+A(r)^{-2}dr^{2}+r^{2}d\th^{2}
+r^{2}\,\sin^2\th\,d\phi^{2}\right)\nonumber \\
&+&a^{2}(r,\x)d\x^2+L^2(r,\x)d\xi^2~.\label{6smetric} \eea The
solutions are summarized in Table 2~\cite{CuadrosMelgar:2008kn}.
\begin{table}[here]
\begin{center}
\begin{tabular}{|c|c|c|c|c|}
\hline $A^2(r)$ & $F(\chi)$ & $L(\chi)$ & $-\Lambda_6$ &
Constraints \& $T^{(B)}$\\ \hline
$1+\frac{r^2}{L_4^2}-\frac{\zeta}{r}$ & $\cosh\left( \frac{\chi}{2\sqrt{3\alpha}}\right)$ & $\forall L(\chi)$  & $\frac{5}{12\a}$ & $\a=\frac{L_4^2}{12}$, \\
&&&&$T^{\chi}_\chi = T^{\xi}_\xi = - \frac{6\a \zeta^2}{r^6 F(\chi)^4}$\\
$1+\frac{r^2}{L_4^2}-\frac{\zeta}{r}$ & $\cosh\left( \frac{\chi}{2\sqrt{3\alpha}}\right)$ & $2\sqrt{3\alpha}\beta. $ & $\frac{5}{12\a}$ & $\a=\frac{L_4^2}{12}$, \\
&&$.\sinh\left( \frac{\chi}{2\sqrt{3\alpha}}\right)$
&&$T^{\chi}_\chi = T^{\xi}_\xi = - \frac{6\a \zeta^2}{r^6 F(\chi)^4}$ \\
$1+\frac{r^2}{L_4^2}-\frac{\zeta}{r}$ & $\pm 1$ & $\frac{\b}{\g} \, \sinh{(\g\,\x)}$ & $\frac{6}{L_4^2} \left(1-\frac{2\a}{L_4^2}\right)$ & $\g = \frac{1}{L_4}\,\sqrt{\frac{1-\frac{L_4^2}{4\a}}{1-\frac{L_4^2}{12\a}}}$, \\
&&&&$T^{\chi}_\chi = T^{\xi}_\xi = - \frac{6\a \zeta^2}{r^6}$\\
$1+\frac{r^2}{L_4^2}-\frac{\zeta}{r}$ & $\pm 1$ & $\frac{\b}{\g} \chi \, \sinh{\g}$ & $\frac{6}{L_4^2} \left(1-\frac{2\a}{L_4^2}\right)$ & $\g = \frac{1}{L_4}\,\sqrt{\frac{1-\frac{L_4^2}{4\a}}{1-\frac{L_4^2}{12\a}}}$, \\
&&&&$T^{\chi}_\chi = T^{\xi}_\xi = - \frac{6\a \zeta^2}{r^6}$, \\
&&&&$T^{br}= \frac{3(4\a-L_4^2)}{L_4^2}$\\
\hline
\end{tabular}
\end{center}
\caption{Black String-Like Solutions in Six-Dimensional
Braneworlds of Codimension-2}\label{table2}
\end{table}

\section{The r$\hat{o}$le of the Gauss-Bonnet Term}

In codimension-2 braneworlds there is a relation connecting the
Gauss-Bonnet term projected on the brane with the components of
the bulk energy-momentum tensor corresponding to the extra
dimensions~\cite{Papantonopoulos:2005ma}. In six dimensions it
reads\footnote{A similar relation  involving the Gauss-Bonnet term
was presented in~\cite{Molina:2008kh} in a different context.} \be
-\frac{1}{2}\,R^{(4)} - \frac{1}{2} \a  \left( R^{(4)\,2} -
4R^{(4)\,2}_{\m\n} +R^{(4)\,2}_{\m\n\k\la} \right) =
\frac{1}{M^4_6}\,T^{(B)\,\chi}_{\chi} -\frac{\La_6}{M^4_6}\,.
\label{6DBulkrr1} \ee All bulk solutions have to satisfy this
relation which acts as a consistency relation.  For the
Schwarzschild-AdS solution of the form $A(r)^2$ appearing in the
above table the square of the Riemann tensor reads \be
 R_{\mu\nu\kappa\lambda}^2=\frac{192\zeta^{2}e^{\frac{4\chi}{L_4}}}{(1+e^{\frac{2\chi}{L_4}})^{4}r^{6}}
+\frac{60}{L_4^{4}}~,\label{riemanntensor} \ee while the Ricci
scalar and Ricci tensor are constants. Therefore, for the relation
(\ref{6DBulkrr1}) to be satisfied the bulk energy-momentum tensor
$T^{(B)\,\chi}_{\chi}|_0$ has to scale as $1/r^{6}$ with the right
coefficients. This is actually what happens considering the result
appearing in the table.  Thus, the presence of the Gauss-Bonnet
term in the bulk, which acts as a source term because of its
divergenceless nature, dictates the form of matter that must be
introduced in the bulk in order to sustain a black hole on the
brane. It is interesting to observe that this "holographic matter"
does not depend explicitly on the extra dimension but only through
the warp function $F(\chi)$ which at large $\chi$ goes to zero. On
the other hand on the brane, in the infrared limit we recover the
conventional four-dimensional gravity, while in the ultaviolet
limit we have strong gravity effects modifying the
four-dimensional gravity in a non-trivial way \footnote{Black hole
solutions in codimension-2 braneworlds were also recently
discussed in~\cite{Charmousis:2008bt}.}.

\section{Conclusions}
 We discussed black holes localization on a thin
brane of codimension-2 and their extension into an  AdS bulk. To
reproduce gravity on the brane, we introduced a  Gauss-Bonnet term
in the bulk and an induced gravity term on the brane. We showed
that black holes can be localized on the conical brane, in five
dimensions  a BTZ black hole while in six dimensions a
Schwarzschild-AdS, and these black hole solutions can be extended
into the bulk with a warp function.
 Consistency of the  bulk
equations requires a fine-tuned relation between the Gauss-Bonnet
coupling constant and the length of the  AdS space. The use of
this fine-tuning gives to the non-singular horizon the shape of a
throat up to the horizon of the AdS space with no other curvature
singularities except the brane string-like singularity.

The presence of the  Gauss-Bonnet term is important in our
considerations. It allows the existence of black string solutions
in five dimensions and in six dimensions it specifies the form of
matter which is needed  in the bulk in order to sustain a black
hole on the brane.

We have not discussed the issue of stability of the solutions we
found. The presence of the GB term in the bulk renders the problem
difficult to tackle.

\section*{Acknowledgments}

This talk is based on the work done in collaboration with B.
Cuadros-Melgar, M. Tsoukalas and V. Zamarias. We thank the
organizers and especially Carlos Romero for their kind invitation.
This work was supported by the NTUA research program PEVE07.


\end{document}